\documentclass[showpacs,prl,twocolumn,aps]{revtex4}

\pdfoutput=1

\usepackage{hyperref}
\usepackage{amsmath}
\usepackage{amssymb}
\usepackage{latexsym}
\usepackage{amsfonts}
\usepackage{epsfig}
\usepackage{psfrag}
\usepackage{graphicx}

\def\hhref#1{\href{http://arxiv.org/abs/#1}{arXiv:#1}} 

\usepackage{color}

\newcommand{\bea}{\begin{eqnarray}}
\newcommand{\ea}{\end{eqnarray}}
\newcommand{\eea}{\end{eqnarray}}

\begin{document}

\title{Ramsey Fringes and Time-domain Multiple-Slit Interference from Vacuum}

\author{Eric~Akkermans$^{1}$ and  Gerald~V.~Dunne$^{2}$}
\affiliation{$^{1}$ Department of Physics, Technion Israel Institute of Technology,
  32000 Haifa, Israel   \\ $^{2}$Department  of Physics, University of Connecticut, Storrs, CT 06269-3046, USA}

\begin{abstract}
Sequences of alternating-sign time-dependent electric field pulses  lead to coherent interference effects in Schwinger vacuum pair production, producing a  Ramsey interferometer, an all-optical time-domain realization of the multiple-slit interference effect, directly from the quantum vacuum. The interference, obeying fermionic quantum statistics, is manifest in the momentum dependence of the number of produced electrons and positrons along the linearly polarized electric field. The central value grows like $N^2$ for $N$ pulses [i.e., $N$ "slits"], and the functional form is well-described by a coherent multiple-slit expression. This behavior is generic for many driven quantum systems.
\end{abstract}



\maketitle


Double-slit 
 experiments form a cornerstone of interferometry in optics and in quantum mechanics.
The 
double-slit 
equivalent in the time domain constitutes Ramsey interferometry \cite{cohen}, and  has been widely studied in 
atomic systems \cite{dalibard,paulus,remetter,mansten}.  
Here we propose a new realization of Ramsey interferometry using the Schwinger effect, namely the non-perturbative production of electron-positron pairs when an external electric field is applied to  the quantum electrodynamical (QED) vacuum \cite{he,schwinger}. The analogy between  double-slit interference and the Schwinger effect 
 was suggested in \cite{Hebenstreit:2009km}, and a spatial realization of an all-optical double-slit experiment using  vacuum polarization effects has been proposed \cite{king}.  A multiple-slit analogy has also been made for finite  plane-wave pulses in stimulated laser pair production \cite{Heinzl:2010vg}.
The elusive Schwinger effect has attracted recent renewed interest, prompted by the possibility of experimental realization in ultra-intense laser field systems \cite{tajima,dunne-eli}. It has been realized that the "Schwinger limit"  laser intensity of $4\times 10^{29} W/{\rm cm}^2$ is not necessarily a strict limit, and might be lowered by several orders of magnitude by  manipulation of the form of the laser pulses \cite{Bulanov:2010ei,Schutzhold:2008pz,DiPiazza:2009py,Dunne:2009gi,Monin:2009aj}.
Here we propose a temporal pulse sequence set-up that acts as a Ramsey interferometer and leads, for the number of pairs created, to an  $N^2$ enhancement  factor  for $N$ pulses, due to coherent quantum interference. Our description relies on a general quantitative method which applies to a broad range of similar interference phenomena for quantum fields of different quantum statistics, driven by time-dependent perturbations.
Interference phenomena are familiar from strong-field atomic and molecular physics,  in the theory of atomic ionization \cite{popov-review}, and form the basis for the interpretation of photoionization spectra as time-domain realizations of the double-slit experiment  \cite{paulus}, for pulses having maximal carrier phase offset.
 Thus, similar ideas apply directly to a wide variety of physical systems involving  time-dependent tunneling \cite{KeskiVakkuri:1996gn},  Landau-Zener effect \cite{hanggi,oka,shev},  driven atomic systems \cite{scully}, chemical reactions \cite{miller,batista}, Hawking radiation \cite{Parikh:1999mf}, cosmological particle production \cite{Parker:1968mv}, heavy ion collisions \cite{greiner,dima}, and the dynamical Casimir effect \cite{reynaud}.

Consider the QED vacuum subject to a linearly polarized  time-dependent electric field $\vec{E}=(0, 0, E(t))$, with vector potential $\vec{A}=(0, 0, A(t))$, and $E(t)=-\dot{A}(t)$. For such an applied field, spatial momentum is a good quantum number, so we  decompose the spinor quantum field operators into modes labelled by their spatial momenta. 
A Bogoliubov transformation from the initial time-independent basis of fermionic particle/antiparticle creation and annihilation operators, $a_{\bf k}$ and $b_{-{\bf k}}^\dagger$,  to a time-dependent basis, $\tilde{a}_{\bf k}(t)$ and $\tilde{b}_{-{\bf k}}^\dagger(t)$, is \cite{popov,kluger,blaschke}:
\begin{eqnarray}
\begin{pmatrix}
\tilde{a}_{\bf k}(t)\cr
\tilde{b}_{-{\bf k}}^\dagger(t)
\end{pmatrix}
=\begin{pmatrix}
\alpha_{\bf k}(t) & -\beta_{\bf k}^*(t)\cr
\beta_{\bf k}(t) & \alpha_{\bf k}^*(t)
\end{pmatrix}
\begin{pmatrix}
a_{\bf k}\cr
b_{-{\bf k}}^\dagger
\end{pmatrix} 
\end{eqnarray}
The fermionic anti-commutation relations are preserved by the unitarity condition, $|\alpha_{\bf{k}}(t)|^2+|\beta_{\bf{k}}(t)|^2=1$, and  
the time evolution is: 
\begin{eqnarray}
\dot{\alpha}_{\bf k} &=& \Omega_{\bf k}(t) \, e^{2i\int^t{\mathcal E}_{\bf k} (t') dt'} \, \beta_{\bf k}(t) \\ \nonumber 
\dot{\beta}_{\bf k} &=& -  \Omega_{\bf k}(t) \, e^{-2i\int^t{\mathcal E}_{\bf k} (t') dt'} \, \alpha_{\bf k}(t)
\label{eq1}
\end{eqnarray}
where 
\begin{eqnarray}
{\mathcal E}^2_{\bf{k}}(t) &=& m^2 + k_{\perp}^2 + (k- A(t))^2  \\ \nonumber
\Omega_{\bf k}(t) &=& E(t) \sqrt{m^2+k_\perp^2}/(2 {\mathcal E}^2_{\bf k}(t)) \, \ .
\label{eq2}
\end{eqnarray}
with the notation ${\bf k} = ({\bf k_\perp}, k)$.  
It is useful to re-express the time evolution (\ref{eq1}) as a two-level problem. For each mode ${\bf k}$, define a two-level system by $c^0_{\bf k}=\alpha_{\bf k}\,e^{-i\int^t {\mathcal E}_{\bf k}}$, $c^p_{\bf k}=\beta_{\bf k}\,e^{i\int^t {\mathcal E}_{\bf k}}$, with time evolution:
\begin{eqnarray}
i {d \over dt} \begin{pmatrix}
{c}^0_{\bf k}\\
{c}^p_{\bf k}
\end{pmatrix}=
\begin{pmatrix}
{\mathcal E}_{\bf k}(t) & i\Omega_{\bf k}(t)\\
-i\Omega_{\bf k}(t) &-{\mathcal E}_{\bf k}(t)
\end{pmatrix}
\begin{pmatrix}
c^0_{\bf k}\\
c^p_{\bf k}
\end{pmatrix}
\label{rabi}
\end{eqnarray}
The off-diagonal matrix elements $\Omega_{\bf k}(t)$ are proportional to the electric field $E(t)$ and can be interpreted as a Rabi frequency. Note that the energies ${\mathcal E}_{\bf k}(t)$ depend also on the field and all matrix elements depend parametrically on ${\bf k}$.

\begin{figure}[htb]
\includegraphics[scale=0.25]{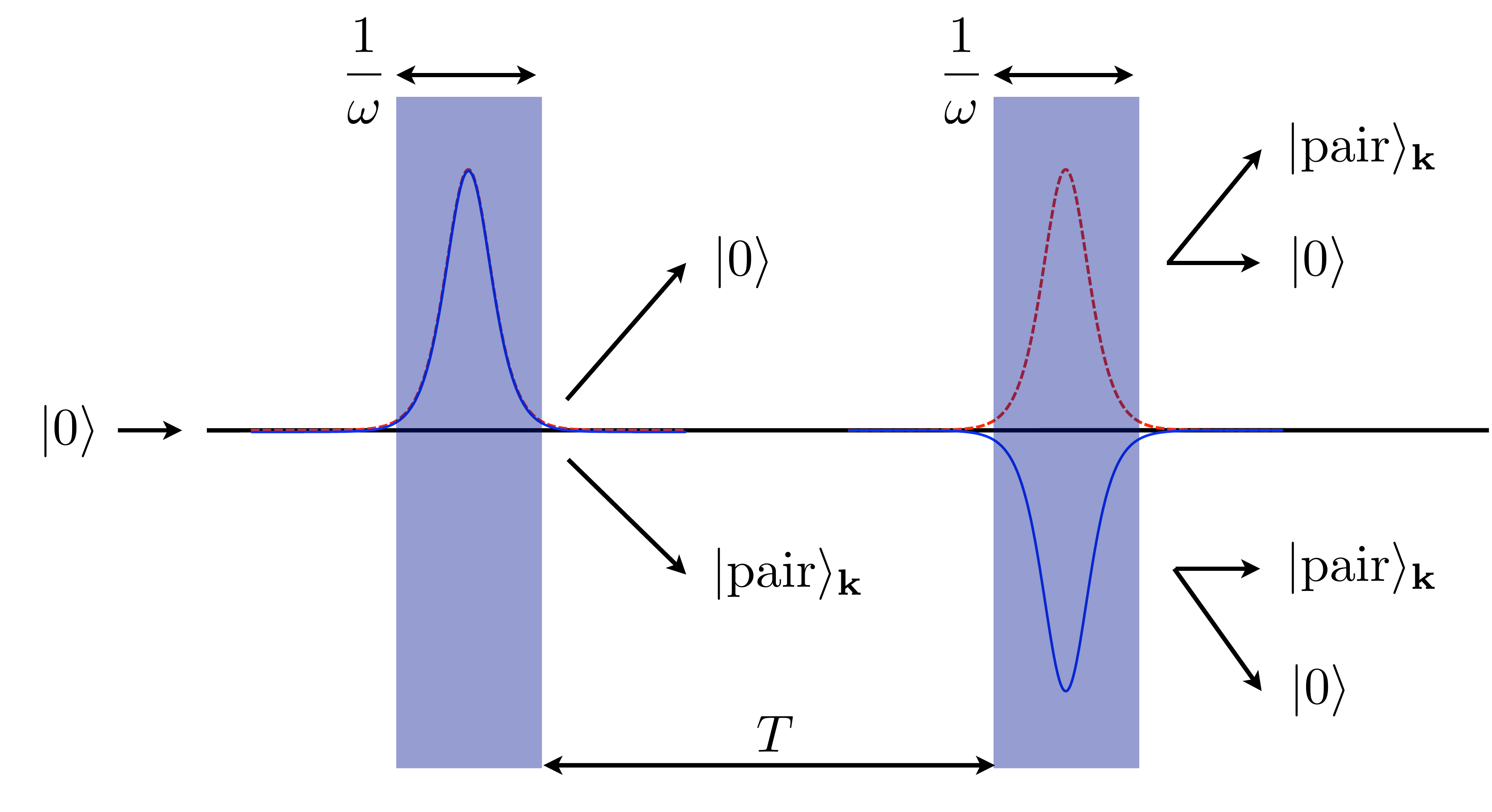}
\caption{Two-pulse Ramsey interferometer. A sequence of two identical electric field pulses of width $1/\omega$, separated by time delay $T$ is applied to the vacuum. We consider the symmetric and antisymmetric situations where the pulses have the same or opposite signs.}
\label{f1}
\end{figure}

The physical quantity we wish to evaluate is the expectation value  ${\mathcal N}_{\bf k}$ of the number of pairs produced from vacuum into the momentum mode ${\bf k}$. It is given by
\begin{equation}
{\mathcal N}_{\bf k}=|\beta_{\bf k}(t=+\infty)|^2 = |c^p_{\bf k}(t=+\infty)|^2
\label{eq4}
\end{equation}
 Time evolution through a single pulse is described by an $S$-matrix written as a rotation characterized by an angle $\phi_{\bf k}$, so that ${\mathcal N}_{\bf k}=\sin^2\phi_{\bf k}$. For two successive pulses, as shown in Fig.\ref{f1}, there are two amplitudes, $A_1=e^{i\theta_{\bf k}}\cos\phi_2\sin\phi_1$ and $A_2=e^{-i\theta_{\bf k}}\cos\phi_1\sin\phi_2$, for producing a pair with momentum ${\bf k}$, where $2\theta_{\bf k}$ is a phase accumulated between the two pulses. For two identical pulses of opposite sign we have $\phi_1=- \phi_2$, and quantum interference leads to:
\begin{eqnarray}
{\mathcal N}_{\bf k}^{\text{2-pulse}} = |A_1 + A_2 |^2 &\approx& 4\sin^2 \theta_{\bf k}  \,\, {\mathcal N}_{\bf k}^{\text{1-pulse}}
\label{twopair}
\end{eqnarray}
assuming ${\mathcal N}_{\bf k}\ll 1$.
An explicit expression for the interference angle $\theta_{\bf k}$ is obtained below (see (\ref{theta})).

\begin{figure}[htb]
\includegraphics[scale=0.5]{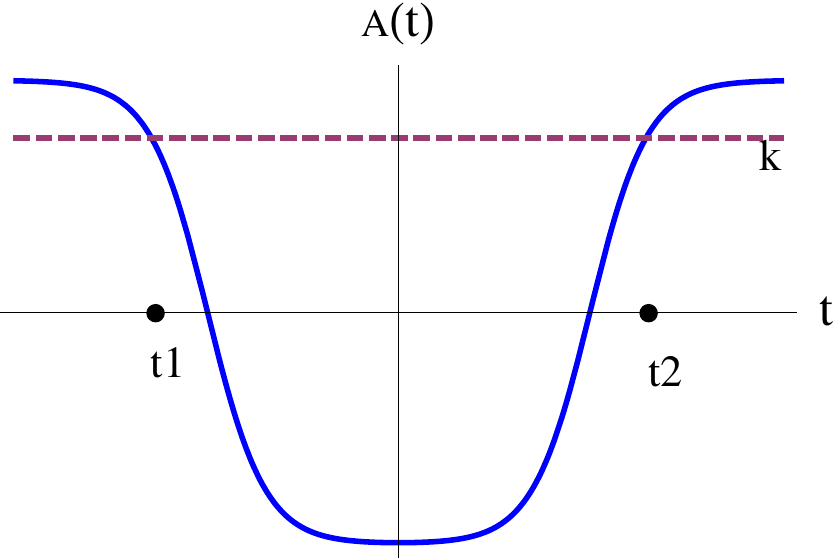}
\includegraphics[scale=0.50]{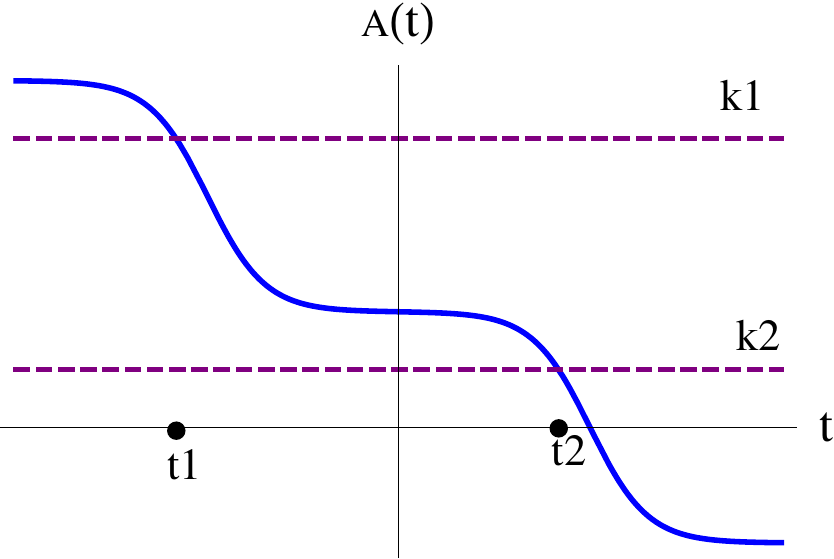}
\caption{Plots of the gauge field $A(t)$. 
The first plot describes the situation of two electric pulses in the antisymmetric configuration. It has two avoided crossings, corresponding to  the same $k$ value, and interference occurs. The second plot, with a monotonic $A(t)$ and correspondingly two identical electric pulses (symmetric configuration), has crossings at different $k$, so no interference takes place.}
\label{f2}
\end{figure}

A qualitative physical understanding of this quantum interference can be given in terms of avoided level-crossings between the instantaneous eigenvalues, $\lambda_\pm=\pm \sqrt{{\mathcal E}_{\bf k}^2+\Omega_{\bf k}^2}$, from (\ref{rabi}). The maximum pair production occurs for the level-crossings at which ${\mathcal E}_{\bf k}\approx 0$ and the Rabi frequency $\Omega_{\bf k}$  is correspondingly large. 
For example, for a constant electric field, $A(t)=-E \, t$, there is an avoided crossing at $t_0=-\frac{k}{eE}\pm i\frac{m}{eE}$. The imaginary part leads to the exponential behavior of the pair number, ${\mathcal N}_{\bf k} \sim \exp\left[-m^2\pi/(eE)\right]$, in analogy with the Landau-Zener argument. The real part, $\mbox{Re}\, t_0 = -\frac{k}{eE}$, depends on the momentum $k$. It indicates the existence of an avoided crossing and creation of pairs of momentum $k$ only.  For two successive electric field pulses, we must distinguish between the  symmetric and antisymmetric configurations of the two pulses as displayed in Fig. \ref{f1}. As shown in Fig. \ref{f2},  the antisymmetric configuration allows for  two distinct level-crossings for the same momentum $k$, and therefore we have interference. 
On the other hand, for the symmetric configuration, we cannot have two different level-crossings for the same momentum $k$, so there is no interference. 

Building on this qualitative description, we now give a precise quantitative treatment. The time-evolution in (\ref{rabi}) can be converted to a   Riccati equation for the ratio $R_{\bf k}=\beta_{\bf k}/\alpha_{\bf k} = \left( {c}^p_{\bf k} / {c}^0_{\bf k} \right) e^{-2i \int^t {\mathcal E}_{\bf k}(t') dt'} $:
\begin{eqnarray}
\dot{R}_{\bf k}
&=&-\Omega_{\bf k} \left(e^{-2i\int^t {\mathcal E}_{\bf k}(t') dt'} + R_{\bf k}^2 \, e^{2i\int^t {\mathcal E}_{\bf k}(t') dt'} \right)
\label{rdot2}
\end{eqnarray}
The amplitude $R_{\bf k}$ is a convenient quantity since, according to (\ref{eq4}), the corresponding probability taken at $t = + \infty$ is, for ${\mathcal N}_{\bf k}\ll 1$,
\begin{equation}
|R_{\bf k}(\infty)|^2=\left| \beta_{\bf k}(\infty)/\alpha_{\bf k}(\infty)\right |^2 \approx {\mathcal N}_{\bf k} \, .
\end{equation}
This relation allows to describe  ${\mathcal N}_{\bf k}$ as the reflection probability
of an associated time-domain scattering problem \cite{brezin,popov}.
While (\ref{rdot2}) can be solved numerically,  
deeper physical insight is gained from a semiclassical approximation \cite{pechukas}. A similar-style analysis for Landau-Zener  is given in \cite{wittig}. The turning points (namely the avoided crossings) obtained for $\mathcal E_{\bf k}(t)=0$, lie in the complex $t$-plane, and since $A(t)$ is real, they occur in complex conjugate pairs. 
Then,  the semiclassical amplitude $R_{\bf k}$ is a sum over contributions from different turning points \cite{dd1}
\begin{eqnarray}
R_{\bf k}(\infty)&\approx& \sum_{t_p} (-1)^p\, e^{i\, \pi/2}\, e^{-2 i \int_{-\infty}^{t_p}{\mathcal E}_{\bf k}(t) \, dt} 
\label{r}
\end{eqnarray}
The  exponents in (\ref{r}) have both real and imaginary parts, so there can be  interference effects for ${\mathcal N}_{\bf k}$, depending on the distribution of turning points.
The alternating sign in (\ref{r}) is from the fermionic statistics.

\begin{figure}[htb]
\includegraphics[scale=0.75]{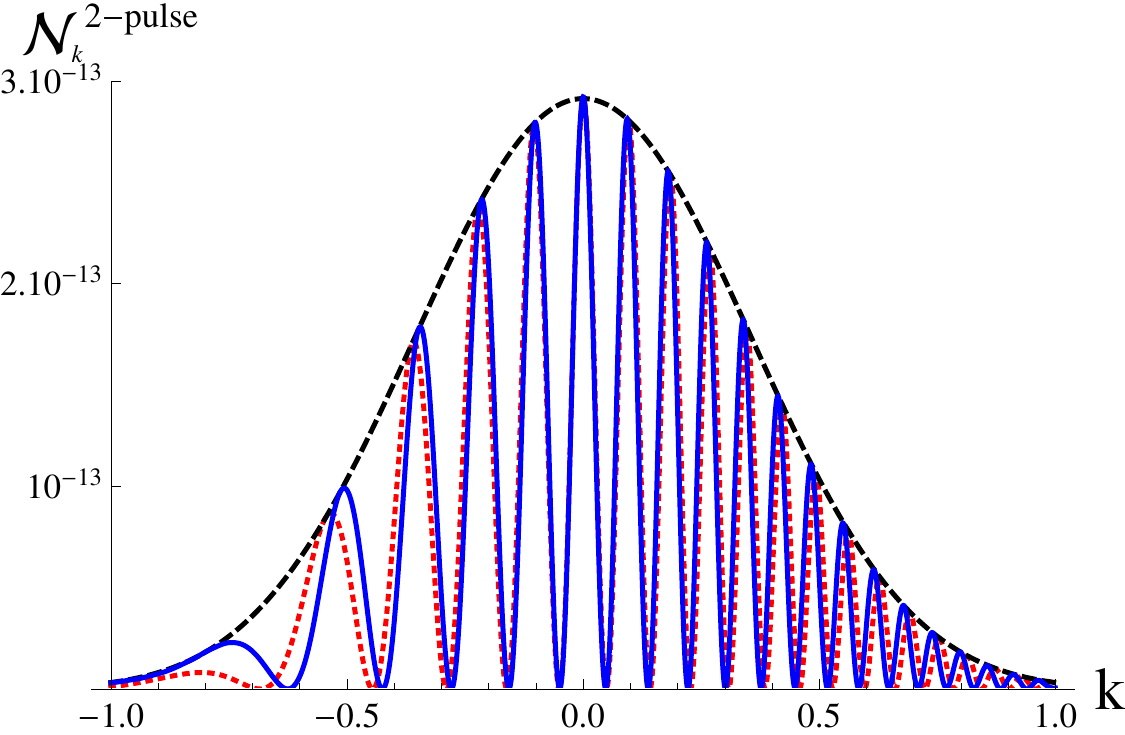}
\caption{Number of pairs created ${\mathcal N}_{\bf k}^{\text{2-pulse}}$, as a function of longitudinal momentum $k$, for the antisymmetric configuration of the two electric pulses. Here $E=.1$, $\omega=.04$, and $T=200.2$,
 all in units where $m=1$.
 The solid [blue] curve is the exact result, the dotted [red] curve is the approximate two-slit expression (\ref{twopair}), and the dashed [black] envelope curve is $2^2$ times the single-slit expression (\ref{onepair}).}
 \label{f3}
\end{figure}

\begin{figure}[htb]
\includegraphics[scale=0.75]{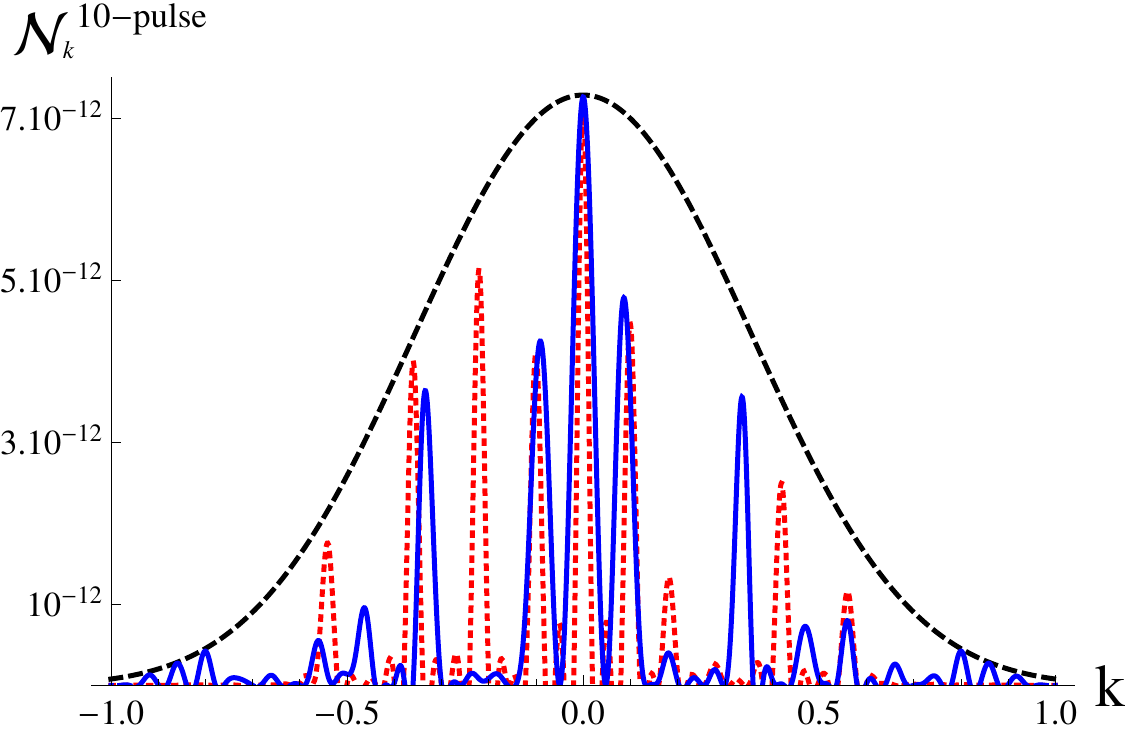}
\caption{As in Fig. 3, now for the $N=10$ antisymmetric configuration of equally spaced pulses. 
The solid [blue] curve is the exact result, the dotted [red] curve is the approximate $N=10$ multiple-slit expression (\ref{multipair}), and the dashed [black] envelope is $10^2$ times the single-slit expression (\ref{onepair}).
}
\label{f5}
\end{figure}

As a quantitative illustration, consider first  a single pulse $E(t)=E\, {\rm sech}^2(\omega \,t)$, using $A(t)=-E/\omega \,\tanh(\omega t)$. 
We set $k_\perp=0$, as the dominant production is along the field direction.
There is an infinite tower of turning points $(t_p, t_p^*)$, given by $\omega t_p ={\rm arctanh}(i\, m-k)+i p \pi$, but the dominant contribution comes clearly from the pair closest to the real $t$ axis. There is no interference, and the number of pairs created in momentum $k$ is well-described by the familiar expression 
\begin{eqnarray}
{\mathcal N}_{\bf k}^{\text{1-pulse}}\approx \exp\left[-2 K_{\bf k}\right]\quad, \quad K_{\bf k}=\left | \int_{t_0}^{t_0^*}{\mathcal E}_{\bf k}(t)\right|
\label{onepair}
\end{eqnarray}
This expression agrees  well with the numerical result, and it is shown in Figs. \ref{f3} and \ref{f5} as a smooth envelope function.

Now consider two such linearly polarized pulses, of opposite sign (antisymmetric configuration), separated by a time delay $T$, namely $E(t)=E\, {\rm sech}^2(\omega \,(t-T/2))-E\, {\rm sech}^2(\omega \,(t+T/2))$, with 
$A(t)=E/\omega[1+ \tanh(\omega (t-T/2))- \tanh(\omega (t+T/2))]$. The turning point structure is now more complicated, but the dominant turning points form two complex conjugate pairs $t_\pm$ and $t_\pm^*$, whose locations are well approximated by 
\begin{eqnarray}
t_\pm (k) =\pm T/2 +\frac{1}{2\omega}\ln\left(\frac{E+\omega(k+i m)}{E-\omega(k+i m)}\right)
\label{tpm}
\end{eqnarray}
These turning points move as functions of  longitudinal momentum $k$, but always form a rectangular array of two complex conjugate pairs with imaginary part of equal magnitude. The integral between the real parts of the different turning points $t_\pm$  yields a quantitative expression for the interference angle $\theta_{\bf k}$ appearing in (\ref{twopair}) \cite{dd1}:
\begin{eqnarray}
 \theta_{\bf k}&=& \int_{Re(t_-)}^{Re(t_+)}{\mathcal E}_{\bf k}(t)dt
 \label{theta}
\end{eqnarray}
These approximate expressions (\ref{twopair}, \ref{theta}) are  shown in Fig. \ref{f3}, in excellent agreement with the exact numerical result. 
Notice the characteristic oscillatory form of a double-slit Ramsey interference pattern, underneath an envelope that is $2^2$ times the single-pulse result in (\ref{onepair}). 

Starting from (\ref{r}), we can also analyze the case of two identical electric pulses in the symmetric configuration. Since the momentum $k$ is fixed, the turning point structure involves now only one pair of (complex conjugate) turning points for the dominant contribution, as for a single pulse, thus leading to no interference. This result is in complete agreement with the more qualitative picture leading to (\ref{twopair}) and explained in Fig. \ref{f2}.

\begin{figure}[htb]
\includegraphics[scale=0.7]{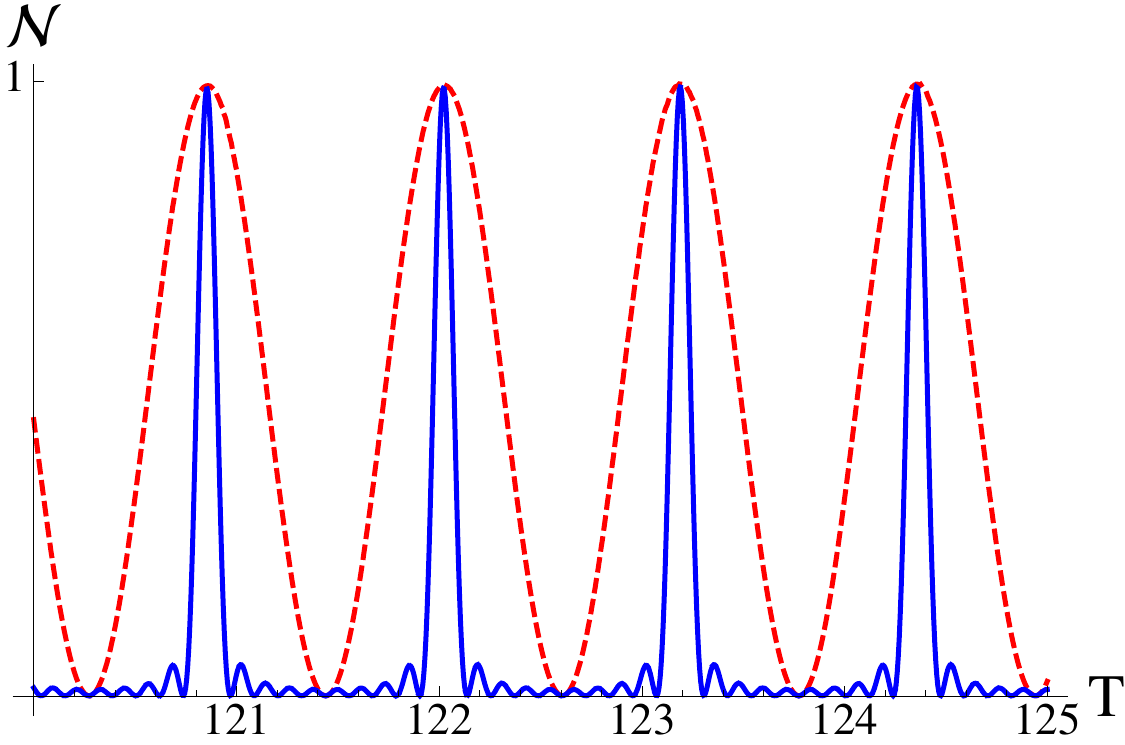}
\caption{The number of pairs created at the central peak value of momentum, normalized by $N^2$ times the single pulse result, as a function of the time delay $T$. The dashed [red] curve  is for $N=2$, and the solid [blue] curve is for $N=10$.}
\label{f4}
\end{figure}

We propose now to generalize the  results (\ref{twopair}, \ref{theta}), for the antisymmetric set-up, to build an interferometer by applying to the QED vacuum a sequence of equally-spaced alternating-sign electric field pulses.  For a fixed momentum mode $k$, the number of pairs created depends on the time-delay $T$ between pulses via the standard Fabry-Perot form, as shown in Fig. \ref{f4}.

For such a field, there are $N$ dominant  complex conjugate pairs of turning points, all equally distant from the real axis, given approximately by the expressions (\ref{tpm}), displaced by steps of $T$ along the real axis. Thus, when the pulses are well-separated compared to their width, $T\gg 1/\omega$, all the $K_{\bf k}^{(p)}$-type  integrals are approximately equal for each set of turning points, and given by $K_{\bf k}$ in (\ref{onepair}). Moreover, the phase integrals $\theta_{\bf k}^{(p, p')}$ between the real parts of the different turning points are approximately integer multiples of the phase integral $\theta_{\bf k}$  for the two-pulse case given in (\ref{theta}). Therefore, the sum over all turning points in (\ref{r}) is coherent, leading to a simple expression for the  number of created pairs,
\begin{eqnarray}
{\mathcal N}_{\bf k}^{\text{N-pulse}}\approx 
\begin{cases}
{\mathcal N}_{\bf k}^{\text{1-pulse}}\, \sin^2\left[ N \theta_{\bf k}\right]/\cos^2\left[ \theta_{\bf k}\right]
 \quad, \quad N\,\,{\rm even}\cr
{\mathcal N}_{\bf k}^{\text{1-pulse}}\, \cos^2\left[ N \theta_{\bf k}\right]/\cos^2\left[\theta_{\bf k}\right]
 \quad, \quad N\,\,{\rm odd}
 \end{cases}
 \label{multipair}
\end{eqnarray}
This result   has the expected form of a Fabry-Perot  interference pattern, with the single-pulse number ${\mathcal N}_{\bf k}^{\text{1-pulse}} = \exp\left[-2 K_{\bf k}\right] $ from (\ref{onepair}) playing the role of the single-slit intensity distribution, modulated by the interference term for $N$ equally spaced slits.

Fig. \ref{f5} shows this approximate multiple-slit result (\ref{multipair}) compared to the numerical result for the $N=10$ antisymmetric configuration of  pulses. The first observation is that the envelope does indeed behave as $N^2$ times the single-pulse profile, behavior characteristic of multiple-slit interference, resulting in a 100-fold increase of the central peak for the ten-slit  configuration.  Furthermore, we see clearly  the narrowing of the central peaks, another feature of multiple-slit interference.
Beyond these  qualitative comments, the quantitative agreement between the semiclassical result (\ref{multipair}) and numerics is also surprisingly  good, especially for the central peaks.

To conclude, in this paper we have described a Ramsey multiple-time-slit interference effect for pairs created from the QED vacuum. We have shown that interference occurs for a sequence of alternating sign pulses of the electric field and we have proposed a qualitative description based on a study of avoided crossings in a two-level system. We have presented a quantitative semi-classical description which gives approximate results in excellent agreement with the exact numerical solutions. The resulting Ramsey interference leads to 
a coherent enhancement, which may 
be viewed as another route towards the Schwinger effect. 
The physical description in terms of quantum interference and avoided-level-crossings is versatile, and suggests that it would be worthwhile studying other more complex pulse sequences,  such as periodic, quasi-periodic or disordered, that might lead to even stronger  (exponential) localization of modes \cite{anderson}. While our QED discussion here was for fermions, both the ideas and analysis generalize straightforwardly to bosons, suggesting potential applications to driven Bose-Einstein condensates or superfluids.

\bigskip

This work was supported in part by  the DOE  grant DE-FG02-92ER40716 and by the Israel Science Foundation grant 924/09 .


\begin{thebibliography}{999}


\bibitem{cohen} C. Cohen-Tannoudji and D. Guery-Odelin, {\it Advances in atomic physics: an overview}, (World Scientific, 2011).

 \bibitem{dalibard}
 P.~Szriftgiser, D.~Gu\'ery-Odelin, M.~Arndt and J.~Dalibard,
 ``Atomic Wave Diffraction and Interference using Temporal Slits'',
 Phys.\ Rev.\ Lett.\ {\bf 77}, 4 (1996).

  \bibitem{paulus}
   F.~Lindner et al,
 ``Attosecond Double-Slit Experiment'',
  Phys.\ Rev.\ Lett.\ {\bf 95}, 040401 (2005).
G. G. Paulus and G. Stania,
``Attosecond Quantum Stroboscope'',
ChemPhysChem {\bf 10}, 875 (2009)

\bibitem{remetter}
T. Remetter et al, 
``Attosecond electron wave packet interferometry'', 
Nature Phys. {\bf 2}, 323  (2006).

\bibitem{mansten}
E. Mansten et al,
``Spectral Signature of Short Attosecond Pulse Trains'',
Phys. Rev. Lett.  {\bf 102}, 083002 (2009).
 
\bibitem{he}
W. Heisenberg and H. Euler,
``Consequences of Dirac's Theory of Positrons'',
Z. Phys. {\bf 98}, 714 (1936).

\bibitem{schwinger}
J.~Schwinger,
``On gauge invariance and vacuum polarization'',
Phys. Rev. {\bf 82} (1951) 664.

\bibitem{Hebenstreit:2009km}
F.~Hebenstreit, R.~Alkofer, G.~V.~Dunne, H.~Gies,
  ``Momentum signatures for Schwinger pair production in short laser pulses
  with sub-cycle structure,''
  Phys.\ Rev.\ Lett.\  {\bf 102}, 150404 (2009)
  [\hhref{0901.2631}].
  ``Quantum statistics effect in Schwinger pair production in short laser pulses,''
   Int. J. Mod. Phys. A {\bf 25}, 2171 (2010)
[\hhref{0910.4457}].


 \bibitem{king}
  B.~King, A. Di Piazza,  C.~H.~Keitel,
  ``A matterless double slit'',
 Nature Photon. {\bf 4}, 92 (2010);
 M.~Marklund,
 ``Fundamental optical physics: Probing the quantum vacuum'',
 Nature Photon. {\bf 4}, 72 (2010).
 
\bibitem{Heinzl:2010vg}
  T.~Heinzl, A.~Ilderton, M.~Marklund,
  ``Finite size effects in stimulated laser pair production,''
  Phys.\ Lett.\  {\bf B692}, 250 (2010).
[\hhref{1002.4018}].


  \bibitem{tajima}
T.~Tajima,
``Prospect for extreme field science'',
Eur. Phys. J. D {\bf 55}, 519 (2009).


\bibitem{dunne-eli}
G.~V.~Dunne,
``New Strong-Field QED Effects at ELI: Nonperturbative Vacuum Pair Production,''
Eur. Phys. J. D {\bf 55}, 327 (2009).
[\hhref{0812.3163}].


  \bibitem{Bulanov:2010ei}
  S.~S.~Bulanov, V.~D.~Mur, N.~B.~Narozhny, J.~Nees, V.~S.~Popov,
 ``Multiple colliding electromagnetic pulses: a way to lower the threshold of
  e+e- pair production from vacuum,''
  Phys.\ Rev.\ Lett.\  {\bf 104}, 220404 (2010)
  [\hhref{1003.2623}].

  \bibitem{Schutzhold:2008pz}
  R.~Sch\"utzhold, H.~Gies, G.~Dunne,
  ``Dynamically assisted Schwinger mechanism,''
  Phys.\ Rev.\ Lett.\  {\bf 101}, 130404 (2008)
  [\hhref{0807.0754}].


 \bibitem{DiPiazza:2009py}
  A.~Di Piazza, E.~Lotstedt, A.~I.~Milstein, C.~H.~Keitel,
  ``Barrier control in tunneling $e^+ e^-$ photoproduction,''
  Phys.\ Rev.\ Lett.\  {\bf 103}, 170403 (2009)
  [\hhref{0906.0726}].
  

    \bibitem{Dunne:2009gi}
  G.~V.~Dunne, H.~Gies, R.~Sch\"utzhold,
  ``Catalysis of Schwinger Vacuum Pair Production,''
  Phys.\ Rev.\  D {\bf 80}, 111301 (2009)
  [\hhref{0908.0948}].


  \bibitem{Monin:2009aj}
  A.~Monin, M.~Voloshin,
``Photon-stimulated production of electron-positron pairs in electric field,''
  Phys.\ Rev.\  D {\bf 81}, 025001 (2010)
  [\hhref{0910.4762}].


\bibitem{popov-review} An excellent review is: 
V. S. Popov,
``Tunnel and multiphoton ionization of atoms and ions in a strong laser field (Keldysh theory)'',
Phys. Usp. {\bf 47}, 855 (2004), Phys. Atom. Nucl. {\bf 68}, 686 (2005).


  \bibitem{KeskiVakkuri:1996gn}
  E.~Keski-Vakkuri, P.~Kraus,
  ``Tunneling in a Time Dependent Setting,''
  Phys.\ Rev.\  D {\bf 54}, 7407 (1996)
  [\hhref{hep-th/9604151}].
  

  \bibitem{hanggi}
D. Zueco, P. H\"anggi, and S. Kohler,
``Landau-Zener tunnelling in dissipative circuit QED'',
New J. Phys. {\bf 10}, 115012 (2008)  [\hhref{0807.1748}].

   \bibitem{oka}
T. Oka and H. Aoki,
``Nonequilibrium Quantum Breakdown in a Strongly Correlated Electron System'', 
\hhref{0803.0422v1}, 
in {\it 
Quantum and Semi-classical Percolation and Breakdown in Disordered Solids}, 
Lect. Notes  Phys., Vol. {\bf 762}, A. K. Sen et al (Eds), (Springer, 2009).


\bibitem{shev}
S. N. Shevchenko, S. Ashhab, F. Nori,
``Landau-Zener-Stuckelberg interferometry'',
Phys. Rept. {\bf 492}, 1 (2010).

 \bibitem{scully}
 H. Li,  V. A. Sautenkov, Y. V. Rostovtsev, M. M. Kash, P. M. Anisimov, G. R. Welch,  M. O. Scully,
 ``Carrier-Envelope Phase Effect on Atomic Excitation by Few-Cycle rf Pulses'',
 Phys.\ Rev.\ Lett.\ {\bf 104}, 103001 (2010);
 P. K. Jha, Y. V. Rostovtsev, H. Li1, V. A. Sautenkov,  M. O. Scully,
``Experimental observation of carrier-envelope-phase effects by multicycle pulses'',
 Phys.\ Rev.\ A {\bf 83}, 033404 (2011).

\bibitem{miller}
W. H. Miller, 
``Semiclassical treatment of multiple turning-point problems -- phase shift and eigenvalues'',
J. Chem. Phys. {\bf 48}, 1651 (1968).

\bibitem{batista}
R. Saha, V. Batista,
``Tunneling under Coherent Control by Sequences of Unitary Pulses'',
J. Phys. Chem. B, {\bf 115}, 5234 (2011).

\bibitem{Parikh:1999mf}
  M.~Parikh, F.~Wilczek,
  ``Hawking radiation as tunneling,''
  Phys.\ Rev.\ Lett.\  {\bf 85}, 5042 (2000)
  [\hhref{hep-th/9907001}].


\bibitem{Parker:1968mv}
  L.~Parker,
  ``Particle creation in expanding universes,''
  Phys.\ Rev.\ Lett.\  {\bf 21}, 562-564 (1968).
  
  \bibitem{greiner}
W.~Greiner, B.~M\"uller and J.~Rafelski,
{\it Quantum Electrodynamics Of Strong Fields}, (Springer, Berlin, 1985).


 \bibitem{dima}
  D.~Kharzeev, E.~Levin and K.~Tuchin,
  ``Multi-particle production and thermalization in high-energy QCD,''
  Phys.\ Rev.\  C {\bf 75}, 044903 (2007)
  [\hhref{hep-ph/0602063}].
  
 \bibitem{reynaud}
 M-T.~Jaekel  and S.~Reynaud,
`` Movement and Fluctuations of the Vacuum'',
Rep. Prog. Phys. {\bf 60}, 863 (1997)
 [\hhref{quant-ph/9706035}];
 V.~V.~Dodonov,
  ``Current status of the dynamical Casimir effect,''
  Phys.\ Scripta {\bf 82}, 038105 (2010).


\bibitem{popov}
V.~S.~Popov,
``Pair Production in a Variable External Field (Quasiclassical approximation)'',
Sov. Phys. JETP {\bf 34}, 709 (1972);
``Pair production in a variable and homogeneous electric field as an oscillator problem''.
Sov. Phys. JETP {\bf 35}, 659 (1972);
M.~S.~Marinov and V.~S.~Popov,
  ``Electron-Positron Pair Creation From Vacuum Induced By Variable Electric Field,''
  Fortsch.\ Phys.\  {\bf 25}, 373 (1977). For a general presentation, see N.D. Birrell and P.C.W. Davies, {\it Quantum fields in curved space}, (Cambridge, 1982).



\bibitem{kluger}
Y. Kluger, J. M. Eisenberg, B. Svetitsky, F. Cooper and E. Mottola, ``Pair production in a strong electric field'',  Phys. Rev. Lett. {\bf 67}, 2427 (1991); 
``Fermion Pair Production In A Strong Electric Field'',  Phys. Rev. D {\bf 45}, 4659 (1992); 
Y. Kluger, E. Mottola and J. M. Eisenberg, ``The quantum Vlasov equation and its Markov limit'',
 Phys. Rev. D {\bf 58}, 125015 (1998) [\hhref{hep-ph/9803372}].


\bibitem{blaschke}
S. M. Schmidt et al
``Relativistic quantum kinetic equation of the Vlasov type for systems with internal degrees of freedom'', Int. J. Mod. Phys.
E {\bf 7}, 709 (1998);
 D. B. Blaschke et al
``Dynamical Schwinger effect and high-intensity lasers: realising nonperturbative QED'', 
Eur. Phys. J. D {\bf 55}, 341 (2009).

   \bibitem{brezin}
E.~Br\'ezin and C.~Itzykson,
``Pair Production In Vacuum By An Alternating Field,''
Phys.\ Rev.\ D {\bf 2}, 1191 (1970).

  \bibitem{pechukas}
J.~P.~Davis and P.~Pechukas,
``Nonadiabatic transitions induced by a time-dependent 
Hamiltonian in the semiclassical/adiabatic limit: The two- 
state case'',
J.\ Chem.\ Phys. {\bf 64}, 3129 (1976).

\bibitem{wittig}
C. Wittig, 
``The Landau-Zener formula'',
 J. Phys. Chem. B {\bf 109}, 8428 (2005).



\bibitem{dd1}
  C.~K.~Dumlu and G.~V.~Dunne,
  ``The Stokes Phenomenon and Schwinger Vacuum Pair Production in
 Time-Dependent Laser Pulses,''
  Phys.\ Rev.\ Lett.\  {\bf 104}, 250402 (2010)
  [\hhref{1004.2509}];
 ``Interference Effects in Schwinger Vacuum Pair Production for Time-Dependent Laser Pulses,''
 Phys.\ Rev.\ D {\bf  83}, 065028 (2011)
   [\hhref{1102.2899}].



\bibitem{anderson} P. Erd\"os and R.C. Herndon, 
``Theories of electrons in one-dimensional disordered
systems'',
Adv.  Phys. {\bf 31}, 65 (1982).
 









\end{thebibliography}
\end{document}